\documentclass[aps,amsfonts,nofootinbib,article,twocolumn,showpacs,superscriptaddress]{revtex4}%{revtex4-1}
\usepackage{graphicx}
\usepackage{amsmath}
\usepackage{amsfonts}
\usepackage{amssymb}

\def\openone{\leavevmode\hbox{\small1\kern-3.8pt\normalsize1}}
\def\N{\leavevmode\hbox{ Z \kern-8 pt\normalsize{Z}}}
\def\openone{\leavevmode\hbox{\small1\kern-3.8pt\normalsize1}}
\def\openJ{\leavevmode\hbox{J \kern-9.5pt\normalsize J}}
\def\openS{\leavevmode\hbox{ S \kern-9.3pt\normalsize S}}
\newcommand{\bb}{\begin{equation}}
\newcommand{\ee}{\end{equation}}
\newcommand{\eqb}{\begin{eqnarray}}
\newcommand{\eqf}{\end{eqnarray}}

\usepackage{color}

\begin{document}

\title{Supersymmetric Majorana Quantum Cosmologies}

\author{Sergio A. Hojman}
\email{sergio.hojman@uai.cl}
\affiliation{Departamento de Ciencias, Facultad de Artes Liberales,
Universidad Adolfo Ib\'a\~nez, Santiago 7941169, Chile.}
\affiliation{Facultad de Ingenier\'{\i}a y Ciencias, Universidad Adolfo Ib\'a\~nez, Santiago 7941169, Chile.}
\affiliation{Departamento de F\'{\i}sica, Facultad de Ciencias, Universidad de Chile,
Santiago 7800003, Chile.}
\affiliation{Centro de Recursos Educativos Avanzados,
CREA, Santiago 7500018, Chile.}

\author{Felipe A. Asenjo}
\email{felipe.asenjo@uai.cl}
\affiliation{Facultad de Ingenier\'{\i}a y Ciencias,
Universidad Adolfo Ib\'a\~nez, Santiago 7941169, Chile.}

\begin{abstract}
The Einstein equations for an isotropic and homogeneous Friedmann--Robertson--Walker Universe in the presence of a quintessence scalar field are shown to be described in a compact way, formally identical to the dynamics of a relativistic particle moving on a two--dimensional spacetime. The correct Lagrangian for the system is presented and used to construct a  spinor quantum cosmology theory using Breit's prescription. The theory is supersymmetric when written in  the Majorana representation. The spinor field components interact through a potential that correlates spacetime metric and the quintessence. An exact supersymmetric solution for  $k=0$ case is exhibited. This  quantum cosmology model may be interpreted as a theory of interacting universes.
\end{abstract}

\pacs{04.20.Fy, 04.50.-h, 04.60.Kz, 12.60.Jv, 98.80.Qc}

\maketitle

The quantization of cosmological Universes has been a very active field for decades.  Quantum cosmology field aims to construct a quantum theory for the entire Universe. Several possibilities have been developed (for a comprehensive review, we refer the reader to Refs.~\cite{bojo,death,vargas}).  One way to proceed is to perform a canonical quantization of the classical dynamical equations for the evolution of the  Friedmann--Robertson--Walker (FRW) Universe. This is achieved by replacing the momentum by a derivative operator with respect to the scale factor variable. The final equation is widely  and generically known as the Wheeler--DeWitt equation \cite{dewitt,hawk1}. This equation has been used to study several features  of a quantum Universe \cite{alvarenga, oliveraG,lemos,mone, mone2,vaki, huang,barbour,hawk2}.

In this work, we study a FRW Universe with dark energy (described by  the quintessence  field). We show that there exists a deep correlation between spacetime and the quintessence field which allows to describe the Universe as a relativistic point-like particle. We study the quantum version of this theory, but instead of using the Wheeler--DeWitt equation, we perform a more fundamental kind of quantization {\it \`a la} Dirac. We find that this quantization allows us to obtain a supersymmetric theory for quantum cosmology.

The evolution of the spacetime metric $g_{\mu \nu}(x^\alpha)$ interacting with the quintessence (massless scalar) field $\phi(x^\beta)$, characterized by a potential ${\cal V}(\phi)$, can be studied using the classical Lagrangian density
\begin{equation}
\mathcal{L}=\sqrt{-g}\left[\frac{1}{2}R-\Lambda-\frac{1}{2}g^{\mu \nu} \phi,_\mu
\phi,_\nu + {\cal V}(\phi)\right]\, ,
\end{equation}
where $g$ stands for the determinant of the metric $g_{\mu\nu}$, $R$ is the  scalar curvature and $\Lambda$ is the cosmological constant. We choose $ 8\pi G/c^4=1$ (where $G$ is the gravitational constant and $c$ is the speed of light).
Notice that the quantities related to the quintessence field appear in the Lagrangian density with a sign which is the opposite of the one used to describe scalar fields. It is worth to mention that the quintessence field is described by the same Lagrangian density as the tlaplon field \cite{hrrs} which describes propagating torsion.
The action $S=\int \mathcal{L} d^4 x$ is varied with respect to the metric tensor and with respect to $\phi$ to get the evolution equations.
Let us consider the simplest case of a Universe described by the line element for an isotropic and
homogeneous FRW spacetime \cite{ryden}
\begin{equation}
ds^2=dt^2- a(t)^2\left[\frac{dr^2}{1-kr^2} +r^2(d\theta^2+
\sin^2\theta d\phi^2)\right]\, ,
\end{equation}
with the usual definitions for the scale factor $a(t)$, and the curvature constant $k$ (where $k$ may be $\pm 1$ or $0$). The dynamics of the FRW in the presence of quintessence is obtained by varying the action $S$, getting the two second order equations
\begin{equation}
2\frac{\ddot a}{a}+\left(\frac{\dot a}{a}\right)^2 +\
\frac{k}{a^2} +\frac{1}{2}\ \dot \phi^2-V(\phi)= 0, \label{frwqe1}
\end{equation}
\begin{equation}
\ddot\phi + 3\frac{\dot a}{a} \dot\phi + \frac{d
V(\phi)}{d\phi}=0\, , \label{frwqe2}
\end{equation}
and one first order constraint
\begin{equation}
3\left(\frac{\dot a}{a}\right)^2 +3\frac{k}{a^2}-\left(\frac{1}{2}\dot \phi^2+ V(\phi) \right) =0\, , \label{frwqc}
\end{equation}
where we introduced $V(\phi)={\cal V}(\phi) - \Lambda$. Eqs. \eqref{frwqe1} and  \eqref{frwqc} correspond to Einstein equations, while Eq.~\eqref{frwqe2} is the Klein-Gordon equation for the quintessence field. Another useful equation can be obtained by manipulating Eqs. \eqref{frwqe1} and \eqref{frwqc}, to get
\begin{equation}
\frac{\ddot a}{a}=-\frac{1}{3}(\dot\phi^2-V)\, .
\end{equation}
 This set of equations has  already been studied  taking advantages of its symmetries \cite{tsu,copeland,capozz}.

The Lagrangian which is usually adopted for this system is
\begin{equation}
L= 3 a {\dot a}^2 - 3 k a - a^3\left(\frac{1}{2}\dot\phi^2-V(\phi) \right)\, , \label{lag1}
\end{equation}
and it gives rise to the dynamical equations \eqref{frwqe1} and \eqref{frwqe2}. However, it is important to emphasize that the Lagrangian \eqref{lag1} does
not produce the constraint equation \eqref{frwqc}. This constraint is equivalent to
imposing that the Hamiltonian $H$ associated to $L$ given by \eqref{lag1} vanishes identically, i.e.,
\begin{equation}
H \equiv \frac{\partial L}{\partial \dot a} \dot a + \frac{\partial
L}{\partial \dot \phi} \dot \phi-  L \equiv 0\, .
\end{equation}
 Now we can introduce the remarkable change of variables
\begin{equation}
\rho=\frac{2\sqrt 6}{3} a^{3/2}\, ,\qquad \theta=\frac{3\phi}{2\sqrt 6}\, ,
\end{equation}
which recasts the Lagrangian $L$ in a ``kinetic energy minus potential energy ($T-V$)'' form (disregarding the fact that both $T$ and $V$ have the ``wrong signs''
in the $\theta$ associated terms)
\begin{equation}
L\rightarrow{\bar L}=\frac{1}{2}(\dot \rho^2 - \rho^2\dot \theta^2)-\bar V(\rho,\theta)\, ,\label{lag2}
\end{equation}
with the general  potential
\begin{equation}
\bar V(\rho,\theta)=3 \left(\frac{3}{8}\right)^{{1}/{3}}k\, \rho^{2/3}-\left(\frac{3}{8}\right) \rho^2 V(\theta)\, .
\end{equation}
Note that the Lagrangian ${\bar L}$ defined in \eqref{lag2}, which
describes the evolution of a Friedmann--Robertson--Walker--Quintessence (FRWQ) Universe constituted by geometry (represented either by $\rho$ or by $a$) and quintessence (represented by  $\theta$ or $\phi$), shows that the Universe evolves as a relativistic particle moving on a two dimensional surface under the influence
of the potential $\bar V$, thus geometrically unifying gravity and quintessence (dark energy).
In order to construct a
Lagrangian which in fact gives rise to all three equations
\eqref{frwqe1}, \eqref{frwqe2} and \eqref{frwqc} it is enough to
recall that \cite{h2} Jacobi--Maupertuis and Fermat principles give rise to
identical equations of motion in classical mechanics and geometrical
(ray) optics except for the fact that Fermat principle also produces
a constraint equation. It is worth mentioning that exactly the same
results are reached for the case of a relativistic particle moving on a two
dimensional conformally flat spacetime.
Therefore, the description of the FRWQ system in terms of a Fermat--type Lagrangian is established by defining the relation between
the potential $\bar V(\rho, \theta)$, the (``refractive index'') conformal factor
\begin{equation}
n^2 (\rho,\theta) =   -2 \bar V(\rho, \theta)\, ,
\end{equation}
and the constraint
\begin{equation}
\bar H \equiv \dot \rho \frac{\partial \bar L}{\partial {\dot\rho}}  +
\dot \theta\frac{\partial \bar L}{\partial {\dot \theta} }  - \bar L \equiv 0\, .
\end{equation}
The Fermat--like Lagrangian $L_F$ which gives rise to all three equations
\eqref{frwqe1}, \eqref{frwqe2} and \eqref{frwqc} is
\begin{equation}
L_F=  \sqrt{2 \bar V(\rho, \theta)\left[ \rho^2 \left(\frac{d
\theta}{d\lambda}\right)^2-\left(\frac{d \rho}{d\lambda}\right)^2  \right]}\, ,\label{LF1}
\end{equation}
where $\lambda$ is,
in principle, an arbitrary parameter.
To reproduce the relativistic equations of motion, $\lambda$ is defined by L\"{u}neburg's parameter
choice \cite{lune,h2}
\begin{equation}
n(\rho,\theta)=\sqrt{\rho^2 \left(\frac{d\theta}{d\lambda}\right)^2-\left(\frac{d \rho}{d\lambda}\right)^2}\, .
\end{equation}
It is straightforward to prove that varying the Lagrangian \eqref{LF1} with respect to $\rho$ and $\theta$ one gets Eqs. \eqref{frwqe1}, \eqref{frwqe2} (when rewritten in terms of $\rho$ and $\theta$) while the constraint \eqref{frwqc} arises much in the same way as $P_{\mu} P^{\mu}-m^2=0$ appears in the Lagrangian description of the dynamics of a relativistic particle. This statement may equivalently be related to the (scalar) eikonal equation in geometrical optics which,
in turn, is equivalent to the (zero energy) Hamilton--Jacobi equation in mechanics.

%%%%%%%%%%%%%%%%%

We can go further in the description of the FRWQ dynamics if we rewrite the Lagrangian \eqref{LF1} as
\begin{equation}
L_F=  \sqrt{\bar V(\xi, \theta)\,  e^{2\xi}\left( \dot\theta^2- \dot\xi^2\right)}\, ,\label{LF2}
\end{equation}
where we have introduced the new variable
\begin{equation}
\xi=\ln \rho\, ,
\end{equation}
 and $\dot\theta={d\theta}/{d\lambda}$, $\dot\xi={d\xi}/{d\lambda}$.
Remarkably, the previous Lagrangian is identical to the one of a relativistic particle in a two-dimensional conformally flat spacetime, with metric $g_{\mu\nu}=\Omega^2\eta_{\mu\nu}$ (where $\eta_{\mu\nu}$ is the flat spacetime metric) and the conformal factor is
\begin{equation}\label{functionOmegaa}
\Omega\equiv\sqrt{\bar V}e^\xi=\left[3 \left(\frac{3}{8}\right)^{\frac{1}{3}}k\, e^{8\xi/3}
-\frac{3}{8} V(\theta) e^{4\xi}\right]^{1/2}\, .
\end{equation}
Thus, the Lagrangian \eqref{LF2} is written as
\begin{equation}
L_F=  \sqrt{g_{\mu\nu} \frac{dx^\mu}{d\lambda} \frac{dx^\nu}{d\lambda}}\, ,
\end{equation}
 with $x^0=\theta$ and $x^1=\xi$.

The correspondence of the spacetime FRW geometry and the quintessence (dark energy) field at the Lagrangian level \eqref{LF2} of a relativistic particle, allows us to study the quantum version of this theory. Quantizing the above FRWQ Lagrangian will give rise to a quantum cosmological model. To avoid problems in the quantization procedure, we study the cases where $\bar V >0$, considering static manifolds  only \cite{saa} where there exists a family of spacelike surfaces always orthogonal to a timelike Killing vector. This implies that ${\partial_\theta g_{\mu\nu}}=0$, or
\begin{equation}
\frac{\partial \bar V}{\partial\theta}=0\, .
\end{equation}
Thereby, the original potential $V(\theta)$ is a constant, which is essentially equal to a cosmological constant.
Under this assumption, the associated Noether conservation law ensures that $\dot\theta$ does not change
sign (see, for instance \eqref{frwqe2} for a $\phi$ independent potential) and, therefore, $\theta$ may be used as the evolution (time) variable which is exactly what the variational principle and quantization procedure suggest.
Therefore, notice that the quintessence field acts as a Super-time in this new description where the particle is moving in an ``effective'' two-dimensional conformally flat spacetime.

Classically, it can be rigorously shown \cite{hanson} that the Hamiltonian for the system described by Lagrangian \eqref{LF2} is
\begin{equation}\label{Hclass1}
  H=\sqrt{g_{00}}\sqrt{1-g^{11}\pi^2}=\sqrt{g_{00}}\sqrt{1+\frac{\pi^2}{\Omega^{2}}}\, ,
\end{equation}
where $\pi$ is the canonical momentum, and $\sqrt{g_{00}}=\Omega$.
We use this Hamiltonian to construct the quantum theory for the FRWQ system. In order to avoid factor ordering issues, the quantum Hamiltonian operator $\mathcal H$ may be constructed from its classical analogue \eqref{Hclass1} as
\begin{equation}\label{momentumoperator1}
  \mathcal H=\Omega^{1/2}\sqrt{1+\hat p^2}\, \Omega^{1/2}\, ,
\end{equation}
where $\hat p$ is the momentum operator
\begin{equation}
\hat p=\sqrt{-g^{11}}\hat\pi=-\frac{i}{\Omega}\frac{\partial}{\partial\xi}\, ,
\end{equation}
because to $\hat\pi=-i{\partial}_\xi$. The quantum equation that describes the quantization of the FRWQ system is
\begin{equation}
i\hbar\frac{\partial\Psi}{\partial\theta}=\mathcal H\Psi\, ,
\end{equation}
where $\Psi$ is the wavefunction for the FRWQ system.
In principle, one may ask whether there are ways to construct other Hamiltonian operators that differ from \eqref{momentumoperator1}, giving rise to quantum theories which are not equivalent to the one described by  $i\hbar{\partial_\theta\Psi}=\mathcal H\Psi$ (see, for instance \cite{hojmanmontemayor}). This point is subtle, and the answer is affirmative, however the operator \eqref{momentumoperator1} has the advantage that it reproduces results from quantum field theory in curved spacetimes as we will show below.

%%%%%%%%%%%%%%%%%%%%%%%%%%

The quantum equation $i\hbar{\partial_\theta\Psi}=\mathcal H\Psi$ is only possible due  to the close association between the FRW geometry and the quintessence scalar field at a Fermat-like Lagrangian level. Each quantum cosmology model requires a closed form for the Hamiltonian \eqref{momentumoperator1}, which basically consists in finding a quantization procedure for which the square-root of the Hamiltonian operator \eqref{momentumoperator1} can be explicitly written.

Solving the square-root for a Klein-Gordon spinless relativistic particle \cite{gavrilov}, one can obtain the Wheeler--DeWitt Super-Hamiltonian formalism. This is the most famous description of quantum cosmology.
However, we can choose a different way to construct the square-root using matrices, in a fashion similar to the quantization of a relativistic particle carried out by Dirac.
This idea has been used before to get  a Dirac square-root formulation (departing  from the Wheeler--DeWitt equation) in a general form in the context of quantum cosmology and supergravity \cite{deathhawk,mallett,kimoh,maciasryan,rgrah,ramirez,socomacias,maobresoc,guven}.
Differently, due to the close relation of theory  \eqref{Hclass1} to a point-like particle descrition, in this work we use a more direct way to proceed with the quantization of this system. Breit \cite{breit}  showed that there exists a correspondence between the Dirac and the relativistic pointlike particle Hamiltonians via a  prescription of replacement of the particle velocity and the Dirac matrices (as well as the prescription in Schr\"{o}dinger or Klein-Gordon theories where the energy and  momentum may be replaced by the time and space derivatives). Thus, Breit's prescription implies a classical and geometrical interpretation of the spin.

Following Breit's ideas,  we can obtain the relation $H^2=g_{00}+\pi^2$ from the classical Hamiltonian \eqref{Hclass1}. From that relation we can obtain
\begin{equation}
\frac{\sqrt{g_{00}}}{H}=\sqrt{1-q^2}\, ,
\end{equation}
where we  defined the velocity variable $q={\pi}/{H}$.
Using this variable, the  square-root of the Hamiltonian \eqref{Hclass1} can be written
\begin{equation}
H=\sqrt{g_{00}}\left(\frac{q\pi}{\Omega}+\sqrt{1-q^2}\right)\, .
\end{equation}
Breit's interpretation  \cite{breit} corresponds in the identification of the Dirac matrices as $q\rightarrow \alpha$, and $\sqrt{1-q^2}\rightarrow\beta$.
Breit showed the these identifications are consistent with the postulates of Dirac electron's theory. The implications of this prescription have been investigated with the purpose of understanding the underlying nature of the spin or antiparticles \cite{savasta}.

Following Breit's prescription, the Hamiltonian operator \eqref{momentumoperator1} can be written using Dirac matrices ($\alpha$ and $\beta$) as
\begin{equation}\label{momentumoperatorDirac}
  {\cal H}=\Omega^{1/2} \left(\alpha\cdot\hat p+\beta\right) \Omega^{1/2}\, ,
\end{equation}
where $\alpha$ and $\beta$ are the flat spacetime Dirac matrices, as the curvature is already taken into account in $\Omega$.
Because of our FRQW system is two dimensional, the matrices are two-dimensional as well (as Dirac matrices are two-dimensional in 2 or 3 dimensional spacetimes). Using the operator \eqref{momentumoperatorDirac}, the quantum mechanical equation $i{\partial_\theta\Psi}=\mathcal H\Psi$ reads (with $\hbar=1$)
\begin{equation}\label{diracQQQ}
  i\gamma^0\frac{\partial\Psi}{\partial\theta}+i\gamma^1\left(\frac{\partial}{\partial\xi}+\frac{1}{2}\frac{\partial\ln\Omega}{\partial\xi}\right)\Psi=\Omega\Psi\, .
\end{equation}
where now $\Psi$ is a spinor, $\gamma^0=\beta$ and $\gamma^1=\gamma^0\alpha$. Because of the dimensionality of the FRWQ system, there exist only two $2\times2$ matrices satisfying the algebra $\{\gamma^\mu,\gamma^\nu\}=2 \eta^{\mu\nu}$.

The  strongest feature of  Breit's ideas and  quantization procedure \eqref{momentumoperatorDirac}  is that the Dirac equation \eqref{diracQQQ}  is exactly the same equation obtained from Quantum Field Theory in curved spacetime \cite{birrel} for the Dirac equation in a  two-dimensional conformally flat space. In general, the curved-space Dirac equation is
\begin{equation}
i {e^\mu}_d \gamma^d \left(\partial_\mu+\frac{1}{8}\omega_{ab\mu}[\gamma^a,\gamma^b]\right)\Psi=\Psi\, ,
\end{equation}
where the vierbein ${e^\mu}_a$ are defined through $g_{\mu\nu}={e_\mu}^a {e_\nu}^b \eta_{ab}$ (and ${e_\mu}^a{e^\mu}_b=\delta^a_b$).
The spin connection is defined as
\begin{equation}
{\omega^c}_{b\mu}={e^c}_\nu {e^\nu}_{b,\mu}+{e^c}_\nu{e^\sigma}_b{\Gamma^\nu}_{\sigma\mu}\, ,
\end{equation}
where ${\Gamma^\nu}_{\sigma\mu}$ are the Christoffel symbols.
In a two-dimensional conformally flat spacetime, the zweirbein components are ${e_0}^0=\Omega$, ${e_1}^1=\Omega$,  ${e^0}_0={1}/{\Omega}$, and ${e^1}_1={1}/{\Omega}$. Thus, $\omega_{ab0}\gamma^a\gamma^b=2\omega_{010}\gamma^0\gamma^1$, $\omega_{ab1}\gamma^a\gamma^b=2\omega_{011}\gamma^0\gamma^1$, $\omega_{010}={\Gamma^0}_{10}=\partial_\xi\ln\Omega$, and $\omega_{011}={\Gamma^0}_{11}=0$.
Using these results, we recover Eq.~\eqref{diracQQQ}. The equivalence of these two different methods reinforces the Breit's procedure.

It is necessary to remark that quantum cosmological theories in a  two-dimensional Lorentzian geometry can be derived in other ways different than using a quintessence model. For example, this has been studied in a spinor model for two-dimensional dilaton gravity \cite{referre1}, or  in the spinor
representation of the Wheeler-DeWitt equation \cite{referre2,referre3}.

Furthermore, with Eq.~\eqref{diracQQQ}, the expected value of the scale factor can be readily calculated as
\begin{equation}\label{scalefactorMU}
\left\langle a(\theta)\right\rangle=\frac{1}{|\Psi|^{2}}\int_0^{a} da\,  \Omega\,  \Psi^\dag \Psi\, ,
\end{equation}
evaluating the integrals and $\Psi$ in terms of $a$. Here,  $\Psi^\dag$ is the transpose conjugated of the wavefunction $\Psi$, and the probability density of the Dirac field is
\begin{equation}
|\Psi(\theta)|^2=\int_0^{a} \frac{da}{a}\Omega\, \Psi^\dag \Psi\, .
\end{equation}
We can see that $\left\langle a\right\rangle$ does depend explicity on the coupling between wavefunctions, and also on the Super-time $\theta$.

Now, let us define the new wavefunction $\Phi=\sqrt{\Omega}\Psi \exp(iE\theta)$, which satisfies
\begin{equation}\label{diracQQQ2}
 \left(-i\gamma^1\frac{d}{d\xi}+\Omega\right)\Phi=E\gamma^0\Phi\, ,
\end{equation}
which can be found using Eq.~\eqref{diracQQQ}.
Eq.~\eqref{diracQQQ2} couples the components of the spinor $\Phi$. Defining the operators $P_+$ and $P_-$,  and the spinor $\Phi$ as
\begin{equation}
P_-=\left(\begin{array}{c}
                          1 \\
                          0\\
                        \end{array}
                      \right)\, ,\quad
P_+=\left(\begin{array}{c}
                          0 \\
                          1\\
                        \end{array}
                      \right)\, ,\quad
\Phi=\left(\begin{array}{c}
                          \varphi_- \\
                          \varphi_+\\
                        \end{array}
                      \right)\, ,
\end{equation}
then $P_+^\dag \Phi=\varphi_+$, and  $P_-^\dag \Phi=\varphi_-$. The system \eqref{diracQQQ2} can now be recasted in the two coupled equations
\begin{equation}
Q_\pm\varphi_\pm+W_\pm \varphi_\mp=0\, ,
\end{equation}
 with the operators
\begin{eqnarray}\label{operaTotal}
Q_\pm&=&-i (P_\pm^\dag \gamma^1 P_\pm)\frac{d}{d\xi}+\Omega-E (P_\pm^\dag \gamma^0 P_\pm)\, ,\nonumber\\
W_\pm&=&-i (P_\pm^\dag \gamma^1 P_\mp)\frac{d}{d\xi}-E (P_\pm^\dag \gamma^0 P_\mp)\, .
\end{eqnarray}
Next, let us focus in the different choices for the matrices which satisfy $\{\gamma^\mu,\gamma^\nu\}=2\eta^{\mu\nu}$, giving different types of evolution equations. One interesting posibility is the choice of
\begin{equation}
P_\pm^\dag \gamma^1 P_\mp=0\, .
\end{equation}
 In this case, it will be fixed the form of the matrices to be
\begin{equation}\label{MajoranaMatrices}
\gamma^0=\left(\begin{array}{cc}
                          0 & 1\\
                          1 &0\\
                        \end{array}
                      \right)\, , \qquad
\gamma^1=\left(\begin{array}{cc}
                          i & 0\\
                          0 &-i\\
                        \end{array}
                      \right)\, ,
\end{equation}
also impliying that
\begin{equation}
P_\pm^\dag \gamma^0 P_\pm=0\, ,\qquad P_\pm^\dag \gamma^1 P_\pm=\mp i\, , \qquad P_\pm^\dag \gamma^0 P_\mp=1\, .
\end{equation}

Matrices \eqref{MajoranaMatrices} correspond to the two-dimensional Majorana representation, which implies that $\Phi$ is real.  Besides, the operators \eqref{operaTotal} reduce to
\begin{equation}\label{OperSUSYMAJO}
Q_\pm=\mp\frac{d}{d \xi}+\Omega\, ,\qquad W_\pm=-E\, .
\end{equation}
 Thus, dynamical equations \eqref{diracQQQ2} become
\begin{equation}\label{SUSYM}
\left(\mp\frac{d}{d\xi}+\Omega\right) \varphi_\pm=E\varphi_\mp\, .
\end{equation}
Eqs.~\eqref{SUSYM} correspond a set of supersymmetric equations of quantum mechanics \cite{death,vargas,cooper,crom,cooper2}.
This  Supersymmetric Quantum Cosmology (SSQC) theory can only be obtained in the Majorana picture. This SSQC scheme is different from previous models \cite{death,vargas,death2,tkach,socorro,vargas2,vargas3,graham,mielke,assaoui,maciasca}, as here the SSQC emerges naturally owing to the correspondence of the FRW spacetime and the quintessence.
As Eq.~\eqref{SUSYM} has a supersymmetric  structure, the two spinor components are the super-partner of each other.
 Each wavefunction satisfies
\begin{equation}\label{hamiltonianequationpm}
H_\pm\varphi_\pm={E^2}\varphi_\pm\, ,
\end{equation}
 where the Hamiltonians operators and potentials are
\begin{equation}
H_\pm=-{d_\xi^2}+{\cal W}_\pm\, ,\qquad {\cal W}_\pm=\pm{d_\xi\Omega}+\Omega^2\, ,
\end{equation}
respectively.
 As usual \cite{cooper2}, we can define the Super-Hamiltonian and the supercharge operators as
\begin{equation}
\mathbf{H}=\left(\begin{array}{cc}
                          H_+ & 0\\
                          0 & H_-\\
                        \end{array}
                      \right)\, ,\,\,
Q=\left(\begin{array}{cc}
                          0 & Q_+\\
                          0 & 0\\
                        \end{array}
                      \right)\, ,\,\,
Q^\dag=\left(\begin{array}{cc}
                          0 & 0\\
                          Q_- & 0\\
                        \end{array}
                      \right)\, ,
\end{equation}
repectively, with  operators defined in Eq.~\eqref{OperSUSYMAJO}. Thus, we can find
\begin{equation}
\mathbf{H}\Phi=E^2\Phi\, ,\qquad Q\Phi=EP_-^\dag\Phi\, , \qquad Q^\dag\Phi=EP_+^\dag\Phi\, .
\end{equation}
 The supercharge operators change bosonic (fermionic) states into fermionic (bosonic) ones.
The above operators have the algebra
\begin{equation}
\mathbf{H}=\{Q,Q^\dag\}\, , \quad Q^2=[\mathbf{H},Q]=0=(Q^\dag)^2=[\mathbf{H},Q^\dag]\, .
\end{equation}
The Hamiltonians $H_\pm$ has the same bound state spectra $E_{n+1}^-=E_n^+$ (for $n=0,1,...$), and the bound states eigenfunctions of the Super-Hamiltonian are related simply by $Q_+\varphi_+^n=E_n^+\varphi_-^{n+1}$, and $Q_-\varphi_-^{n+1}=E_n^+\varphi_+^n$.

Besides,  the energy $E_0^-=0$ of the ground state    of $H_-$, has the wavefunction
\begin{equation}\label{phimenosenergycero}
\varphi_-^0(\xi)=\exp\left(-\int^\xi \Omega(\xi')d\xi'\right)\, .
\end{equation}

The Majorana SSQC theory described by Eqs.~\eqref{SUSYM} is general. An exact solution for the flat curvature $k=0$ case can be obtained explicitly. In this case
\begin{equation}
\Omega=\sqrt{\frac{-3V}{8}}e^{2\xi}\, ,
\end{equation}
 provided that we can choose $V<0$. This gives rise to the Morse potential
\begin{equation}
{\cal W}_\pm=\pm 2 e^{-2 x}+e^{-4x}\, ,
\end{equation}
 with the change of variable $x=-\xi+\xi_0$ and $V=-(8/3)\exp(-4\xi_0)$.
 It is well known \cite{genden,cooperkha} that the Morse potential is one of the solvable  ``shape invariant" potentials in Supersymmetry. This kind of potentials fulfills ${\cal W}_+(x,\varepsilon)-{\cal W}_-(x,\varepsilon_i)=C(\varepsilon_i)$, where $\varepsilon$ is a set of parameters, $\varepsilon_i=f(\varepsilon)$, and $C(\varepsilon_i)$ is a constant.
For the shape invariant potentials the bound state spectrum of $H_-$ is completely determined in terms of the sum of constants $E_n^-=\sum_{i=1}^n C(\varepsilon_i)$ \cite{genden,cooperkha}.  In this form, the $n$-state wavefunction associated to $H_-$ is completely determined \cite{dutt} by $\varphi_-^n(\xi,\varepsilon)=Q_+(\xi,\varepsilon)Q_+(\xi,\varepsilon_1)...Q_+(\xi,\varepsilon_{n-1})\varphi_-^0(\xi,\varepsilon_n)$, and hence the $n$-state wavefunction associated to $H_+$ can be found.

 We find an explicit solution for $k=0$, where the solution \eqref{phimenosenergycero} for the wavefunction with vanishing energy becomes simply
\begin{equation}\label{solucionvarphi0}
\varphi_-^0(\xi)=\exp\left(-\frac{1}{4}\sqrt{\frac{-3V}{2}}e^{2\xi}\right)\, .
\end{equation}
Notice that this expression represents a bound state solution because $\varphi_-^0(\xi\rightarrow\infty)= 0$ (where $\xi\rightarrow\infty$ is equivalent to $a\rightarrow\infty$). Also, in general, Eq.~\eqref{hamiltonianequationpm} for $\varphi_-$, reads
\begin{equation}\label{eqk0difMorsevalid}
\frac{d^2 \varphi_-}{dx^2}+\left(E^2+2 e^{-2x}-e^{-4x}\right)\varphi_-=0\, ,
\end{equation}
in terms of $x$. The solutions for the Morse potential are known. They represent a diatomic molecule with a minimum equilibrium distance bound equal to $-\xi_0$.
It is straightforward to prove that in the  region $-\infty<\xi <\infty$ (or $0<a <\infty$), the only possible solution for bound-states is for $E=0$   \cite{cooper2, corderohojman, dahl, filho, donghan, sukumar, flugge,taseli}.

On the other hand, there are no bound-states solutions for $\varphi_+$. This is due to the fact that the potential ${\cal W}_+= 2 e^{-2 x}+e^{-4x}$ which appears in Eq.~\eqref{hamiltonianequationpm} for  $\varphi_+$ is positive everywhere.

For other curvature cases $k=\pm 1$ a deeper analysis is required and this task is left for future works. Nevertheless, the SSQC theory \eqref{SUSYM} is interesting as  its origin is the  form of the Lagrangian \eqref{LF2}, which couples gravity and quintessence (dark energy) field. It is for this reason that the Majorana SSQC evolves in $\theta$ and $\xi$ (and not in $t$ and $a$ as other models in SSQC). As far as we know, the association between variables made in \eqref{LF2} has not been envisaged previously, producing  a SSQC theory different from the Wheeler--DeWitt ones.
 As an aside, we would like to mention that other Quantum Cosmology models are possible in this theory. For example, by choosing
\begin{equation}
P_\pm^\dag \gamma^1 P_\pm=0\, ,
\end{equation}
the matrices are fixed to be in the Dirac representation, implying that
\begin{equation}
P_\pm^\dag \gamma^0 P_\mp=0\, , \quad P_\pm^\dag \gamma^0 P_\pm=\mp 1\, , \quad P_\pm^\dag \gamma^1 P_\mp=\pm 1\, .
\end{equation}
 In this representation, the operators become
\begin{equation}
Q_\pm=\Omega\pm E\, , \qquad W_\pm=\mp i \frac{d}{d \xi}\, .
\end{equation}
 Hence, it is straightforward to see that the Dirac representation does not produce a SSQC theory.

Finally, we stress that the Dirac-like equation \eqref{diracQQQ} describes a  spinor quantum cosmology,  representing a generalization of the Wheeler--DeWitt formalisms.  The spinor components  are coupled, and in comparison with  Dirac theory for  particles, we can argue that the spinor components represent two interacting Universes, represented by the components $\varphi_\pm$. Thus, the spinor could work as a description for a Multiverse. The interaction between the wavefunction of the Universes gives the dynamical evolution of the expected value of the scale factor.

For example, the physical meaning of the solution for $k=0$ shown above can be envisaged as a Multiverse theory of Universes behaving as a diatomic molecule in a bound state with vanishing energy and a  minimum equilibrium distance directly related to the quintessence potential by $(1/4)\ln(-3V/8)$.

 This Multiverse theory is different from the Everett's Many-worlds interpretation of quantum mechanics \cite{everett1,everett2}, as each spinor component could be interpreted using Everett's theory. The Many-worlds interpretation of Multiverses have been largely studied (the literature is vast and some few examples are in Refs.~\cite{weinberg,tegmark,tegmark2,tegmark3,feeney,hall,vilekin,carroll1,carroll2,carroll3}).

The Multiverse interpretation of the wavefunction in Eq.~\eqref{diracQQQ} is not new in quantum cosmology (see for example Refs.~\cite{guth,weinstein, robles, larsen}). However, unlike other theories, the mathematical equivalence between this SSQC theory [Eqs.~\eqref{diracQQQ} or \eqref{diracQQQ2}] and  the Dirac theory for spin relativistic particles, make the Multiverse interpretation  a  compelling issue.

\begin{acknowledgments}

F. A. A. thanks the CONICyT-Chile for partial support through Funding No. 79130002. S. A. H. thanks Rafael Rosende for his enthusiastic and unselfish support.

\end{acknowledgments}

\end{document}